\documentclass[runningheads]{llncs}
\usepackage{graphicx}
\usepackage{color}
\usepackage{xcolor}
\usepackage{amssymb}
\usepackage{amsmath}
\usepackage{float}
\usepackage{environ} 
\usepackage{lipsum} 
\usepackage{breqn}
\usepackage{placeins}
\usepackage{bm}
\usepackage{hyperref}

\NewEnviron{SMALL}{%
    \scalebox{0.89}{$\BODY$} 
}
\NewEnviron{SMALLL}{%
    \scalebox{0.94}{$\BODY$} 
}

\begin{document}
\title{Statistical prediction of extreme events from small datasets
\thanks{A. Racca is supported by the EPSRC-DTP and the Cambridge Commonwealth, European \& International Trust under a Cambridge European Scholarship. L. Magri is supported by the ERC Starting Grant PhyCo  949388.}}
\titlerunning{Statistical prediction of extreme events}
%
\author{Alberto Racca\inst{1} \and
Luca Magri\inst{2,1,3}}
\authorrunning{A. Racca and L. Magri}
%
\institute{Department of Engineering, University of Cambridge, UK
\\
\and
Aeronautics Department, Imperial College London, London, UK 
\\
\and
The Alan Turing Institute, London, UK 
}
\maketitle              
\begin{abstract}
We propose Echo State Networks (ESNs) to predict the statistics of extreme events in a turbulent flow.
We train the ESNs on small datasets that lack information about the extreme events. We asses whether the networks are able to extrapolate from the small imperfect datasets and predict the heavy-tail statistics that describe the events. We find that the networks correctly predict the events and improve the statistics of the system with respect to the training data in almost all cases analysed. 
This opens up new possibilities for the statistical prediction of extreme events in turbulence.

\end{abstract}

\section{Introduction}
Extreme events arise in multiple natural systems, such as oceanic rogue waves, weather events and earthquakes \cite{farazmand2019extreme}. A way to tackle extreme events is by computing their statistics to predict the probability of their occurrence. Because extreme events are typically rare, information about the heavy tail of the distribution that describes the events is seldom available. This hinders the performance of data-driven methods, which struggle to predict the events when extrapolating from imperfect datasets \cite{sapsis2021statistics}. In this work, we assess the capability of a form of reservoir computing, the Echo State Network \cite{lukovsevivcius2012practical}, to predict the statistics of extreme events in a turbulent flow \cite{moehlis2004low}. In particular, we analyse the ability of the networks to improve the prediction of the statistics of the system with respect to the available training data. 
The paper is organised as follow. Section \ref{sec:MFE} introduces the turbulent flow model. Section \ref{sec:ESN} describes the Echo State Network. Section \ref{sec:stats} analyses the statistical prediction of extreme events. We summarize the work and present future developments in section \ref{sec:conc}. 

\begin{figure}[H]
\centering
\includegraphics[width=1.0\textwidth]{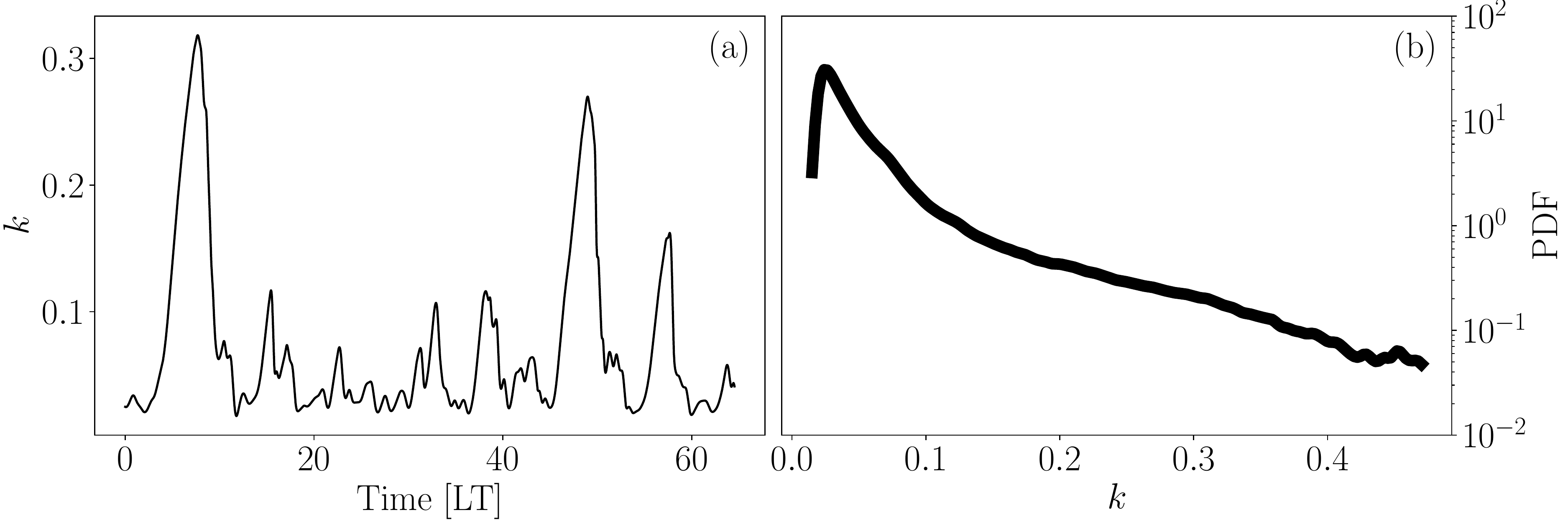}
\includegraphics[width=.66\textwidth]{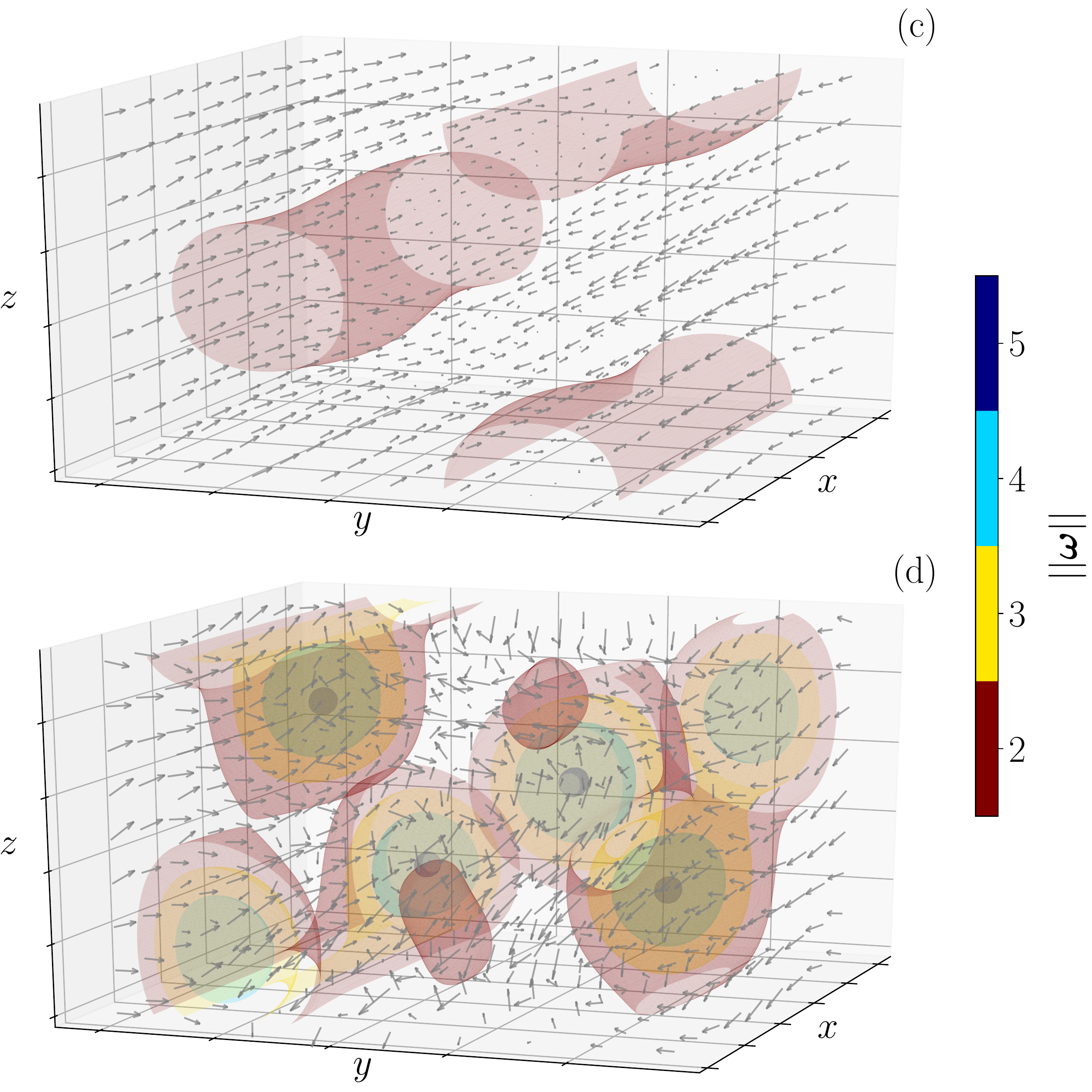}
\caption{One time series of the kinetic energy, (a), and Probability Density Function of the kinetic energy computed from the entire dataset, (b). The time in panel (a) is normalized by the Lyapunov time. Vorticity isosurfaces, $\bm{\omega} = \nabla \times \mathbf{v}$, and velocity flowfield before, (c), and after, (d), an extreme event. The laminar structure, (c), breaks down into vortices, (d).}
\label{fig:k_time_pdf}
\end{figure}


\section{A low-dimensional model for turbulent shear flow}

\label{sec:MFE}

We study a nine-equation model of a shear flow between infinite plates subjected to sinusoidal body forcing \cite{moehlis2004low}. The incompressible Navier-Stokes equations are
\begin{gather}
    \frac{d\mathbf{v}}{dt} = - (\mathbf{v}\cdot\nabla)\mathbf{v} - \nabla p + \frac{1}{\mathrm{Re}}\Delta  \mathbf{v}  + \mathbf{F}(y),
    \label{eq:NS}
\end{gather}
where $\mathbf{v}=(u,v,w)$ is the velocity, $p$ is the pressure, Re is the Reynolds number, $\mathbf{F}(y)=\sqrt{2}\pi^2/(4\mathrm{Re})\sin(\pi y/2)\mathbf{e}_x$ is the body forcing along $x$, $y$ is the direction of the shear between the plates and $z$ is the spanwise direction. 
We solve the flow in the domain $L_x\times L_y\times L_z$, where the boundary conditions are free slip at $y\pm L_y/2$, and periodic at $x=[0;L_x]$ and $z=[0;L_z]$. Here, we set $L_x=4\pi, L_y=2, L_z=2\pi$ and $\mathrm{Re}=400$ \cite{snrinivasan2020}. We project \eqref{eq:NS} on compositions of Fourier modes, $\mathbf{\hat{v}}_i(\mathbf{x})$, so that the velocity is
    $\mathbf{v}(\mathbf{x},t) = \sum_{i=1}^9a_i(t)\mathbf{\hat{v}}_i(\mathbf{x})$ . The projection
generates nine nonlinear ordinary differential equations for the amplitudes, $a_i(t)$, which are the state of the system \cite{moehlis2004low}. The system displays a chaotic transient that converges to the laminar solution $a_1=1,a_2=\dots=a_9=0$. In the turbulent transient, the kinetic energy,
\begin{equation}
    k=0.5\sum_{i=1}^{9}a_i^2,
\end{equation}
shows intermittent large bursts, i.e. extreme events, panel (a) in  Fig. \ref{fig:k_time_pdf}, which generate the heavy tail of the distribution \cite{sapsis2021statistics}, panel (b). In the figure, time is expressed in Lyapunov Times (LT), where a LT is the inverse of the Lyapunov exponent, $\Lambda \simeq 0.0163$. The Lyapunov exponent is the average exponential rate at which arbitrarily close trajectories diverge, which is computed with the QR algorithm \cite{ginelli2007characterizing,huhn2019stability}. Each extreme event is an attempt of the system to reach the laminar solution. During an extreme event, the flow slowly laminarizes, panel (c), but the laminar structure violently breaks down into vortices, panel (d). To study only the transient, we (i) generate $2000$ time series series of length of $4000$ time units through a 4th order Runge-Kutta scheme with $dt=0.25$, (ii) discard all the time series that laminarized, i.e. the ones with $k\geq0.48$, and (iii) use the remaining time series as data. The different time series are obtained by randomly perturbing a fixed initial condition \cite{snrinivasan2020}. 

\begin{figure}[H]
\centering
\includegraphics[width=.8\textwidth]{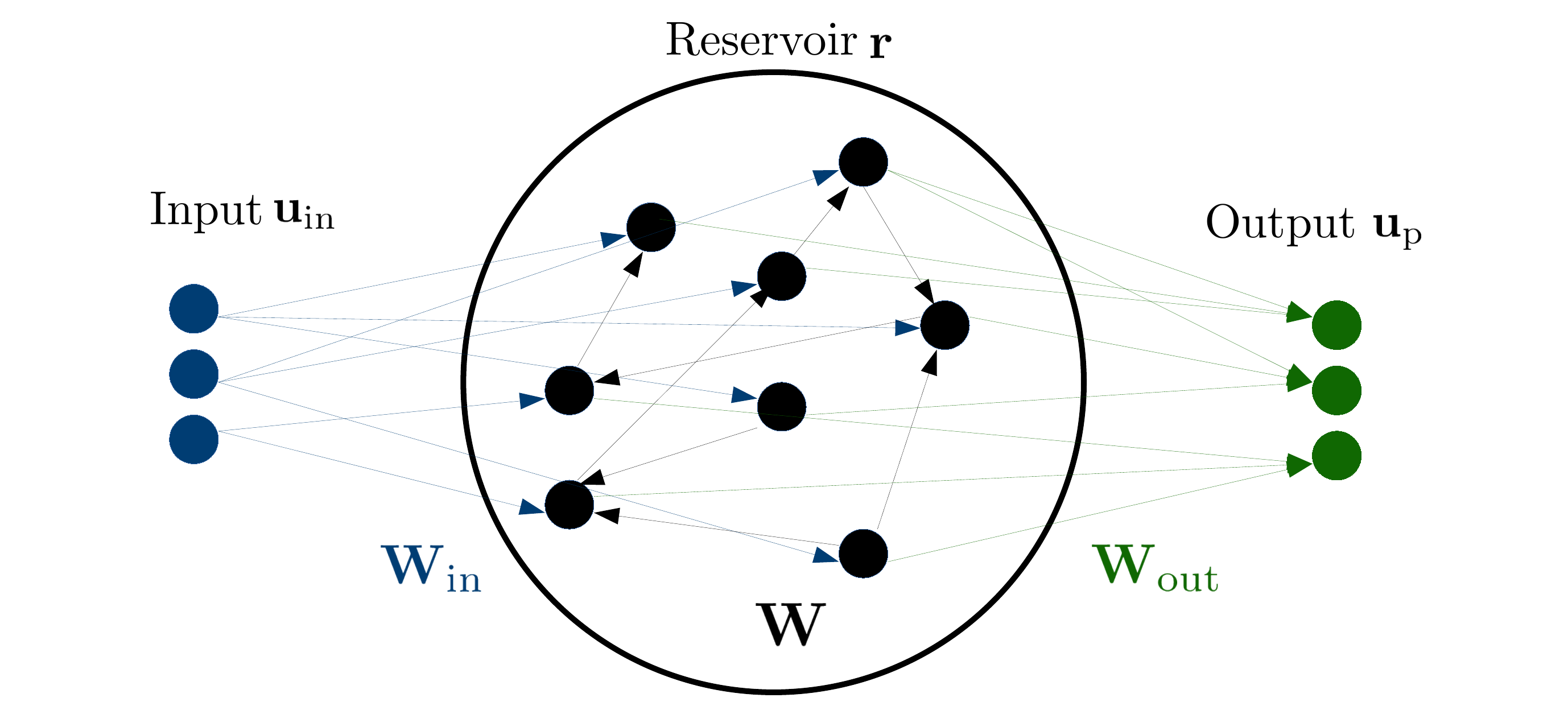}
\caption{Schematic representation of the Echo State Network.}
\label{fig:ESN}
\end{figure}

\section{Echo State Networks}

\label{sec:ESN}

As shown in Fig. \ref{fig:ESN}, in an Echo State Network \cite{lukovsevivcius2012practical}, at the $i$-th time step the high-dimensional reservoir state, $\textbf{r}(t_i)\in\mathbb{R}^{N_r}$, is a function of its previous value and the current input, $\mathbf{u}_{\mathrm{in}}(t_i)\in\mathbb{R}^{N_u}$. The output, $\mathbf{u_{\mathrm{p}}}(t_{i+1})$, which is the predicted state at the next time step, is a linear combination of $\textbf{r}(t_i)$:
\begin{equation}
\label{state_step}
        \textbf{r}(t_i) = \textrm{tanh}\left(\mathbf{W}_{\mathrm{in}}[\mathbf{\Tilde{\mathbf{u}}_{\mathrm{in}}}(t_i);b_{\mathrm{in}}]+
        \mathbf{W}\textbf{r}(t_{i-1})\right); \quad \mathbf{u_{\mathrm{p}}}(t_{i+1}) = \mathbf{W}_{\mathrm{out}}[\mathbf{r}(t_i);1]
\end{equation}
where $\Tilde{(\;\;)}$ indicates normalization by the range component-wise, $\mathbf{W}\in\mathbb{R}^{N_r\times N_r}$ is the state matrix, $\mathbf{W}_{\mathrm{in}}\in\mathbb{R}^{N_r\times (N_u+1)}$ is the input matrix, $\mathbf{W}_{\mathrm{out}}\in\mathbb{R}^{ N_{u}\times(N_r+1)}$ is the output matrix, $b_{\mathrm{in}}$ is the input bias and $[\;;\;]$ indicates vertical concatenation.
$\mathbf{W}_{\mathrm{in}}$ and $\mathbf{W}$ are sparse, randomly generated and fixed. These are constructed in order for the network to satisfy the echo state property \cite{lukovsevivcius2012practical}. The input matrix, $\mathbf{W}_{\mathrm{in}}$, has only one element different from zero per row, which is sampled from a uniform distribution in $[-\sigma_{\mathrm{in}},\sigma_{\mathrm{in}}]$, where $\sigma_{\mathrm{in}}$ is the input scaling. The state matrix, $\textbf{W}$, is an Erdős-Renyi matrix with average connectivity $\langle d\rangle$. This means that each neuron (each row of $\mathbf{W}$) has on average only $\langle d\rangle$ connections (non-zero elements). The value of the non-zero elements is obtained by sampling from an uniform distribution in $[-1,1]$; the entire matrix is then scaled by a multiplication factor to set its spectral radius, $\rho$. 
The only trainable weights are those in the the output matrix, $\mathbf{W}_{\mathrm{out}}$. Thanks to the architecture of the ESN, training the network by minimizing the Mean Square Error (MSE) on $N_{\mathrm{t}} + 1$ points consists of solving the linear system
\begin{equation}
\label{RidgeReg}
    (\mathbf{R}\mathbf{R}^T + \beta \mathbf{I})\mathbf{W}_{\mathrm{out}}^T = \mathbf{R} \mathbf{U}_{\mathrm{d}}^T,
\end{equation}
where $\mathbf{R}\in\mathbb{R}^{(N_r+1)\times N_{\mathrm{t}}}$ and $\mathbf{U}_{\mathrm{d}}\in\mathbb{R}^{N_u\times N_{\mathrm{t}}}$ are the horizontal concatenation of the reservoir states with bias, $[\mathbf{r};1]$, and of the output data, respectively; 
$\mathbf{I}$ is the identity matrix and $\beta$ is the Tikhonov regularization parameter \cite{lukovsevivcius2012practical}. 
\\
The input scaling, $\sigma_{\mathrm{in}}$, spectral radius, $\rho$, and Tikhonov parameter, $\beta$, are selected using Recycle Validation \cite{racca2021robust} to minimize the MSE of the kinetic energy. The Recycle Validation is a recent advance in hyperparameter selection in Recurrent Neural Networks, which is able to exploit the entire dataset while keeping a small computation cost. To minimize the function provided by the validation strategy, we use Bayesian Optimization for $\sigma_{\mathrm{in}}$ and $\rho$ in the interval $[0.1,10]\times[0.1,1]$ seen in logarithmic scale and perform a grid search in each $[\sigma_{\mathrm{in}},\rho]$ point to select $\beta$ from $[10^{-6}, 10^{-9},10^{-12}]$. We set $b_{\mathrm{in}}=0.1$, $d=20$ and add gaussian noise with zero mean and standard deviation, $\sigma_n=0.01\sigma_u$, where $\sigma_u$ is the standard deviation of the data, to the training data \cite{vlachas2020backpropagation}.

\section{Statistical prediction of extreme events}

\label{sec:stats}

We study the capability of the networks to predict the statistics of the system through long-term predictions. Long-term predictions are closed-loop predictions, i.e. predictions where we feed the output of the ESN as an input for the next time step, which lasts several tens of Lyapunov Times. These predictions diverge from the true trajectory due to the chaotic nature of the signal, but remain in the region of phase space of the chaotic transient. In doing so, they replicate the statistics of the true signal. The long-term predictions are generated in the following way: (i) from 500 different starting points in the training set, we generate 500 different time series by letting the ESN evolve each time for 4000 time units ($\simeq 65$LTs); (ii) we discard the laminarized time series and (iii) use the remaining ones to compute the statistics as done for the data, see section \ref{sec:MFE}.
To quantitatively assess the prediction of the statistics, we use the Kantorovich metric \cite{kantorovich1942translocation}, $\mathcal{K}$, also known as Earth mover's distance, and the Mean Logarithmic Error (MLE) with respect to the true Probability Density Function of the kinetic energy, $\mathrm{PDF}_{\mathrm{True}}(k)$,
\begin{gather}
    \mathcal{K} = \int_{-\infty}^{\infty}|\mathrm{CDF}_{\mathrm{True}}(k) - \mathrm{CDF}_j(k)|dk, \\
    \mathrm{MLE} = \sum_{i=1}^{n_{b}}n_b^{-1} |\log_{10}(\mathrm{PDF}_{\mathrm{True}}(k)_i - \log_{10}(\mathrm{PDF}_j(k)_i)|,
\end{gather}
where $\mathrm{CDF}$ is the Cumulative Distribution Function, $j$ indicates the PDF we are comparing with the true data and $n_b$ is the number of bins used in the PDF. When a bin has a value equal to zero and the logarithm is undefined, we saturate the logarithmic error in the bin to be equal to 1.
On the one hand, we use the Kantorovich metric to assess the overall prediction of the PDF of the kinetic energy.
On the other hand, we use the MLE to assess the prediction of the extreme events, as the logarithm highlights the errors in the small values of the tail.

\begin{figure}[H]
\centering
\includegraphics[width=.62\textwidth]{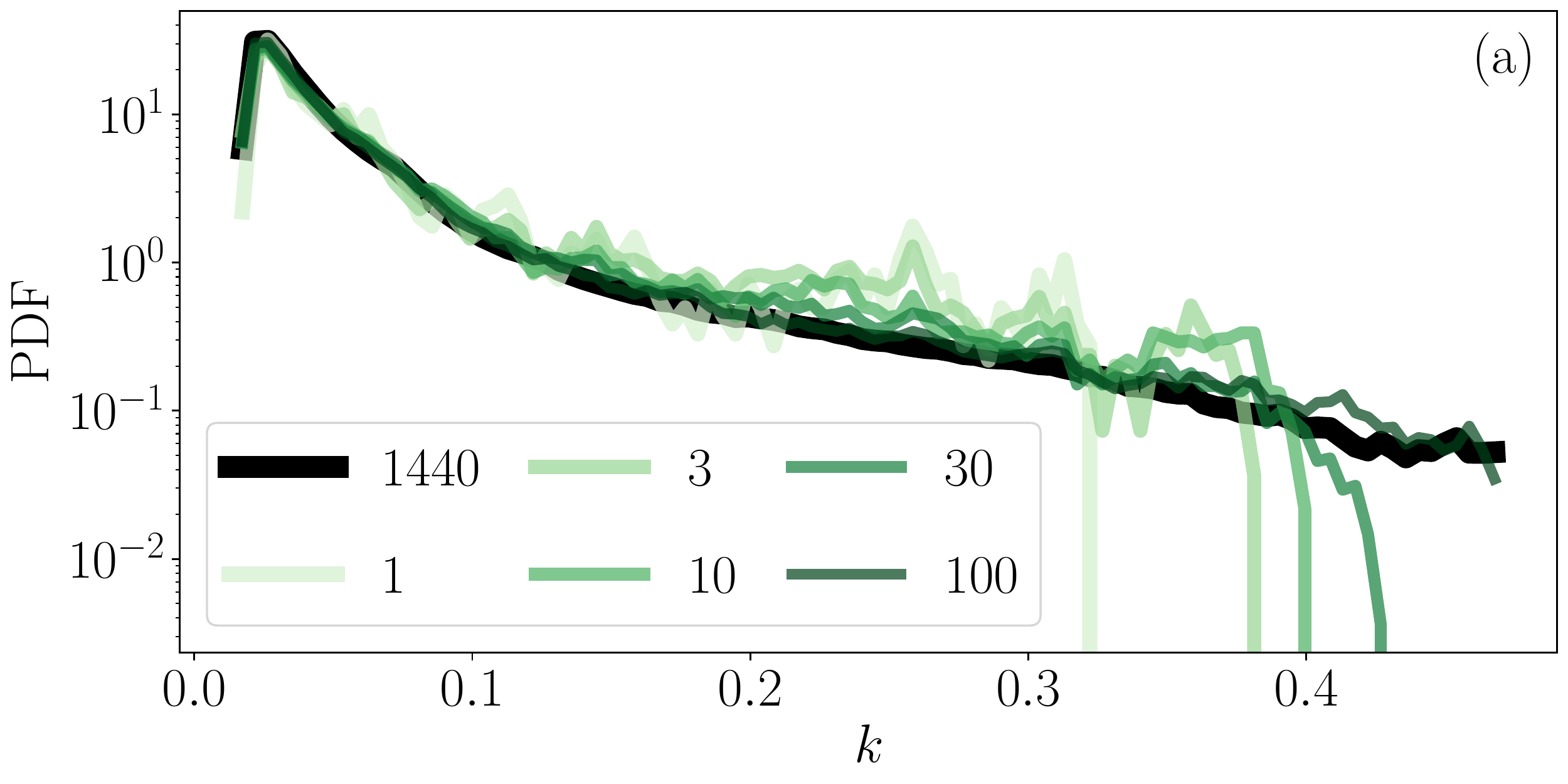}
\includegraphics[width=.93\textwidth]{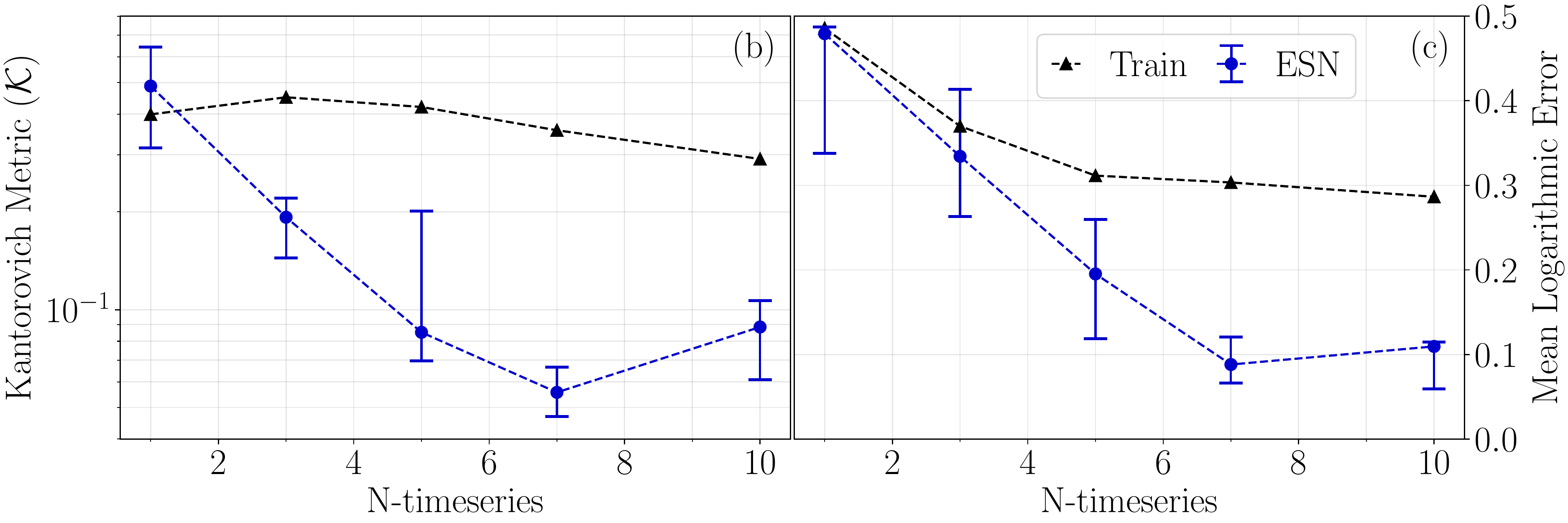}
\caption{PDF of the kinetic energy for different number of time series up to the entire data of 1440 time series, (a). For example, 3 means that the PDF is computed from 3 out of 1440 time series. 25th, 50th and 75th percentiles of Kantorovich Metric, (b), and MLE, (c), for the training set (Train), and the networks (ESN) as a function of the number of time series in the training set (N-timeseries). For example, N-timeseries means that the training set consists of N out 1440 time series.}
\label{fig:small_PDF}
\end{figure}

\noindent In Fig. \ref{fig:small_PDF}, we compare the statistics of the training data and an ensemble of 10 networks of 2000 neurons. We do so because the objective of predicting the statistics is to improve our knowledge, by employing the networks, with respect to the already available knowledge, the training data. Panel (a) shows the PDF of the kinetic energy in the training set for different sizes of the training set, from 1 time series to the entire data (1440). The prediction of the PDF improves with the size of the datasets, and values of the tail up to laminarization are observed only after 100 time series. The unresolved tail due to lack of data is a signature problem of data-driven analysis of extreme events \cite{sapsis2021statistics}. Panels (b)-(c) show the Kantorovich metric and the MLE of the training sets and networks as a function of the training set size. The networks improve the prediction of the PDF with respect to the available data in all figures of merits analyzed, except for one outlier. The MLE of the training set improves more than the Kantorovich metric as the dataset becomes larger. This happens because a small amount of data is needed to accurately describe the peak of the PDF, which affects more the Kantorovich metric, while many time series are needed to describe the tail, which affects more the Mean Logarithmic Error. 
The results indicate that the networks are able to extrapolate from an imperfect dataset and improve the prediction of the overall dynamics of the system.

\begin{figure}[H]
\centering
\includegraphics[width=1.\textwidth]{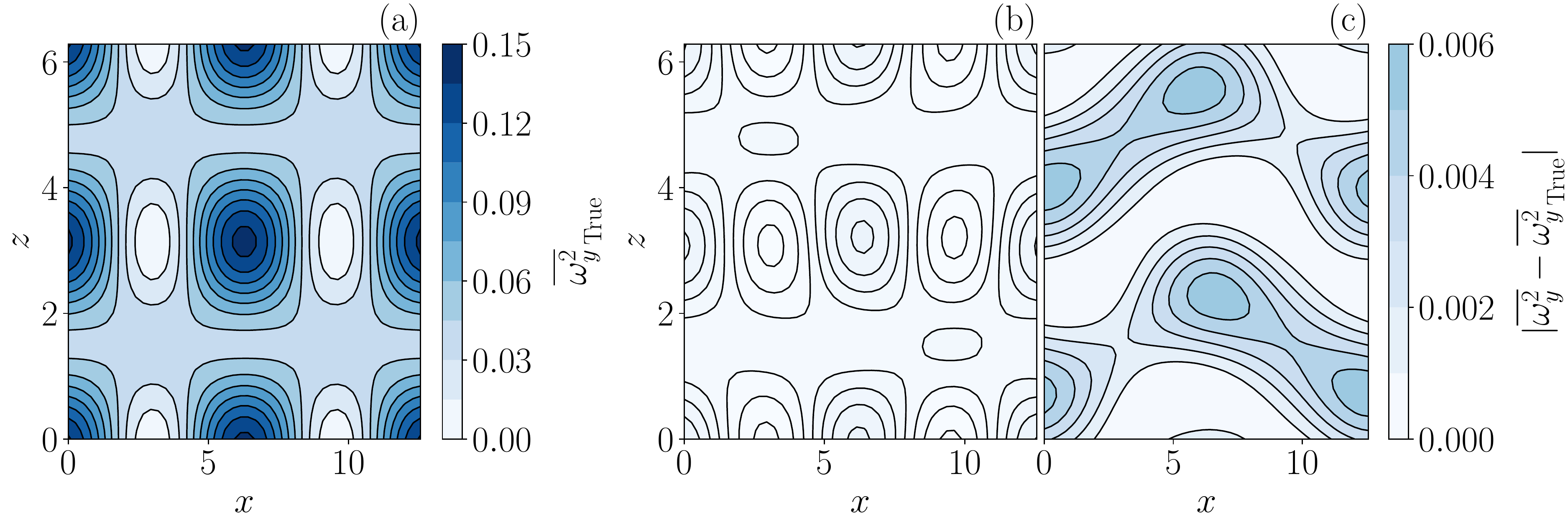}
\caption{Time average of the square of the midplane vorticity, $\omega_y^2(x,0,z)$ for the entire data, (a), error for a 2000 neurons network, (b), and training set, (c).}
\label{fig:v_stats_flows}
\end{figure}

\noindent Fig. \ref{fig:v_stats_flows} shows the statistics of the square of the normal vorticity to the midplane, $\omega_y = \frac{\partial u}{\partial z} - \frac{\partial w}{\partial x}$.
We plot the square of the vorticity, $\omega_y^2$, because the symmetry of the problem causes the time-average of the vorticity to be equal to zero.
Panel (a) shows the flowfield of the time-average, $\overline{(\;)}$, for the entire data, while panels (b) and (c) show the error with respect to (a) for an Echo State Network and the ten time series training set, respectively.  All networks in the ensemble decrease the average error, up to values 7 times smaller than the training data (results not shown). This means that the Echo State Networks are able to extrapolate the statistics of the flowfield in addition to the statistics of the kinetic energy. 
%
\section{Conclusions and future directions}

We propose Echo State Networks to predict the statistics of a reduced-order model of turbulent shear flow that exhibits extreme events. We train fully data-driven ESNs on multiple small datasets and compare the statistics predicted by the networks with the statistics available during training. We find that the networks improve the prediction of the statistics of the kinetic energy and of the vorticity flowfield, sometimes by up to one order of magnitude. This means that the networks are able to extrapolate the statistics of the system when trained on small imperfect datasets.
Future work will consist of extending the present results to higher-dimensional turbulent systems through the combination of Echo State Networks and autoencoders.\\
The code is available on the github repository \href{https://github.com/MagriLab/ESN-MFE}{MagriLab/ESN-MFE}.

\label{sec:conc}
\FloatBarrier

%
%

%

\end{document}